\documentclass[10pt,letterpaper]{article}
\usepackage{opex3}
\usepackage{amssymb}

\usepackage{xcolor}

\begin{document}




\title{Noise in laser speckle correlation and imaging  techniques}

\author{S.E. Skipetrov$^{1}$, J. Peuser$^{2}$, R. Cerbino$^{2,3}$, P. Zakharov$^{4}$, B. Weber$^{5}$ and F. Scheffold$^{2}$ }

\address{$^{1}$Laboratoire de Physique et Mod\'{e}lisation des Milieux Condens\'{e}s, Universit\'{e} Joseph Fourier and
CNRS, 25 rue des Martyrs, 38042 Grenoble, France}
\address{$^{2}$Department of Physics, University of Fribourg, 1700 Fribourg, Switzerland}
\address{$^{3}$Current adress: Department of Chemistry, Biochemistry and Medical Biotechnologies, Universit‡ degli Studi di Milano, I-20133 Milano, Italy}
\address{$^{4}$Solianis Monitoring AG, Leutschenbachstr. 46,
8050 Zurich, Switzerland
\address{$^{5}$Institute for Pharmacology and Toxicology, University of Zurich,
	8091 Zurich, Switzerland}

}

\email{Sergey.Skipetrov@grenoble.cnrs.fr, frank.scheffold@unifr.ch} 


\begin{abstract}
We study the noise of the intensity variance and of the intensity correlation and structure functions measured in light scattering from a random medium in the case when these quantities are obtained by averaging over a finite number $N$ of pixels of a digital camera. We show that the noise scales as $1/N$ in all cases and that it is sensitive to correlations of signals corresponding to adjacent pixels as well as to the effective time averaging (due to the finite sampling time) and spatial averaging (due to the finite pixel size). Our results provide a guide to estimation of noise level in such applications as the multi-speckle dynamic light scattering, time-resolved correlation spectroscopy, speckle visibility spectroscopy, laser speckle imaging etc.
\end{abstract}




\section{Introduction}

The statistical properties of optical speckle patterns resulting from scattering of light in a random medium are largely independent of the nature of the latter \cite{goodman}. Rearrangements of scatterers due to Brownian motion or flow lead to characteristic fluctuations of the speckle intensity with time. Probing these intensity fluctuations is a very common and highly sensitive non-invasive method to study the dynamics of complex fluids and biological systems. One of the earliest applications is \emph{dynamic light scattering} (DLS), also known as photon correlation spectroscopy (PCS), where the fluctuations of the far-field speckle are analyzed \cite{Berne:DLS}. The method is widely used to study polymers in solution or for the sizing of sub-micron particles. The same method applied to turbid media is known as \emph{diffusing-wave spectroscopy} (DWS). In both DLS and DWS, one usually makes use of the ergodicity of intensity fluctuations and replace ensemble averaging by time averaging. Both are typically applied to study fluctuations in the (sub-) millisecond range.  Thus, for a total measurement time of a few minutes, millions of fluctuations are sampled which provides an excellent signal to noise ratio. Using a single photon counter and a digital correlator, nanosecond time resolution is commonly achieved. Even though at the shortest times the signal to noise ratio is limited by photon shot noise \cite{Schatzel83}, in most practical cases this fundamental limitation is not the main concern and other sources of noise (such as sample purity, stability of the experimental setup, etc.) play the main role.

Over the last decade, a wealth of new laser-speckle based experimental techniques have been introduced. Most of them are made possible by recent advances in optical sensor technology. A modern digital CCD (charge-coupled device) or CMOS (complementary metal oxide semi-conductor) sensor contains millions of detectors which allow massive parallel processing of a large number of signals corresponding to intensities of distinct speckle spots. The availability of an area detector has lead to numerous new applications in traditional far-field speckle detection such as multi-speckle dynamic light scattering \cite{Kirsch96,luca2000,harden2000,PineRSI2002,PavelAppop2006}, time-resolved correlation spectroscopy (TRC) \cite{CipTRC2003}, speckle visibility spectroscopy (SVS) \cite{durian,durian:svs2005} and has also enabled new speckle imaging techniques such as the near-field scattering (NFS) \cite{Cerbino2009}, laser speckle imaging (LSI) \cite{briers96:lasca,Pavel2009OE, Baravian2005,Duncan2008} or echo speckle imaging \cite{Zakharov2009, Zakharov2010}.

The obvious advantages of massive parallel detection provided by an image sensor come at a price of specific and often critical limitations. The temporal resolution of digital cameras (typically, in the millisecond range) is inferior to the resolution of a traditional single photon counter such as a photo-multiplier (nanoseconds). In addition, to maximize the signal to noise ratio the experimental setup is usually designed in such a way that the size of an individual speckle spot is comparable to the size of the individual active area (pixel) of the camera. Hence, in contrast to the traditional PCS experiment, a camera pixel is not an ideal point-like detector and, furthermore, fluctuations detected by neighboring pixels are correlated. It would be, of course, desirable to increase the speckle size with respect to the size of the camera pixel (to mimic a point-like detector) and to reduce correlation of signals detected by neighboring pixels (to decrease noise in the statistical properties of speckle patterns determined by averaging over a group of pixels). However, these two objectives are mutually exclusive, at least if one wishes to exploit signals from \textit{all} available pixels of the image sensor.

The limitations of digital cameras as image sensors appear most pronounced in such applications as, for example, LSI. In LSI an image of a sample acquired by the camera is divided in a set of ``meta-pixels'', typically 25--100 pixels each. The spatially resolved information about statistical properties of the sample (an ``image'' in the statistical sense) is obtained by analyzing statistical properties of speckle (mean, variance, etc.) within each meta-pixel separately. Typically, the measurements are made in  backscattering from, for example, a biological tissue, and allow a full-field monitoring of dynamic properties, such as blood flow.  The method is widely used for biomedical studies  \cite{briers:review,dunn:lsi01,weber:imaging,durduran:spat04,dunn:spat05} since it provides access to physiological processes \textit{in vivo} with excellent temporal and spatial resolution. LSI is also becoming increasingly popular in soft material sciences as a probe of heterogenous dynamic properties \cite{Zakharov2010,Erpelding2008}. Obviously, speckle imaging techniques require a tradeoff between spatial resolution and statistical accuracy \cite{Duncan2008}. Due to the rather low number of pixels forming a single meta-pixel (25--100 as already mentioned) the statistical accuracy and noise are of major concern for the data analysis.

The goal of this article is to provide guidelines for a better understanding of the measured quantities and their fluctuations in speckle-based optical techniques for essentially all cases of practical interest. Whenever possible we provide analytical expressions that can be directly applied to the analysis of experimental data. In particular, we analyze the influence of a limited time resolution and provide an approximate treatment of spatial correlations between signals corresponding to neighboring pixels. We start by considering the variance of intensities in a stationary speckle patterns and then extend our analysis to the time correlation function and the intensity structure function of dynamic speckle patterns.

\section{Properties of stationary speckle patterns}
\label{static}

\begin{figure}[t]
\centering{
\includegraphics[width=12cm,angle=0]{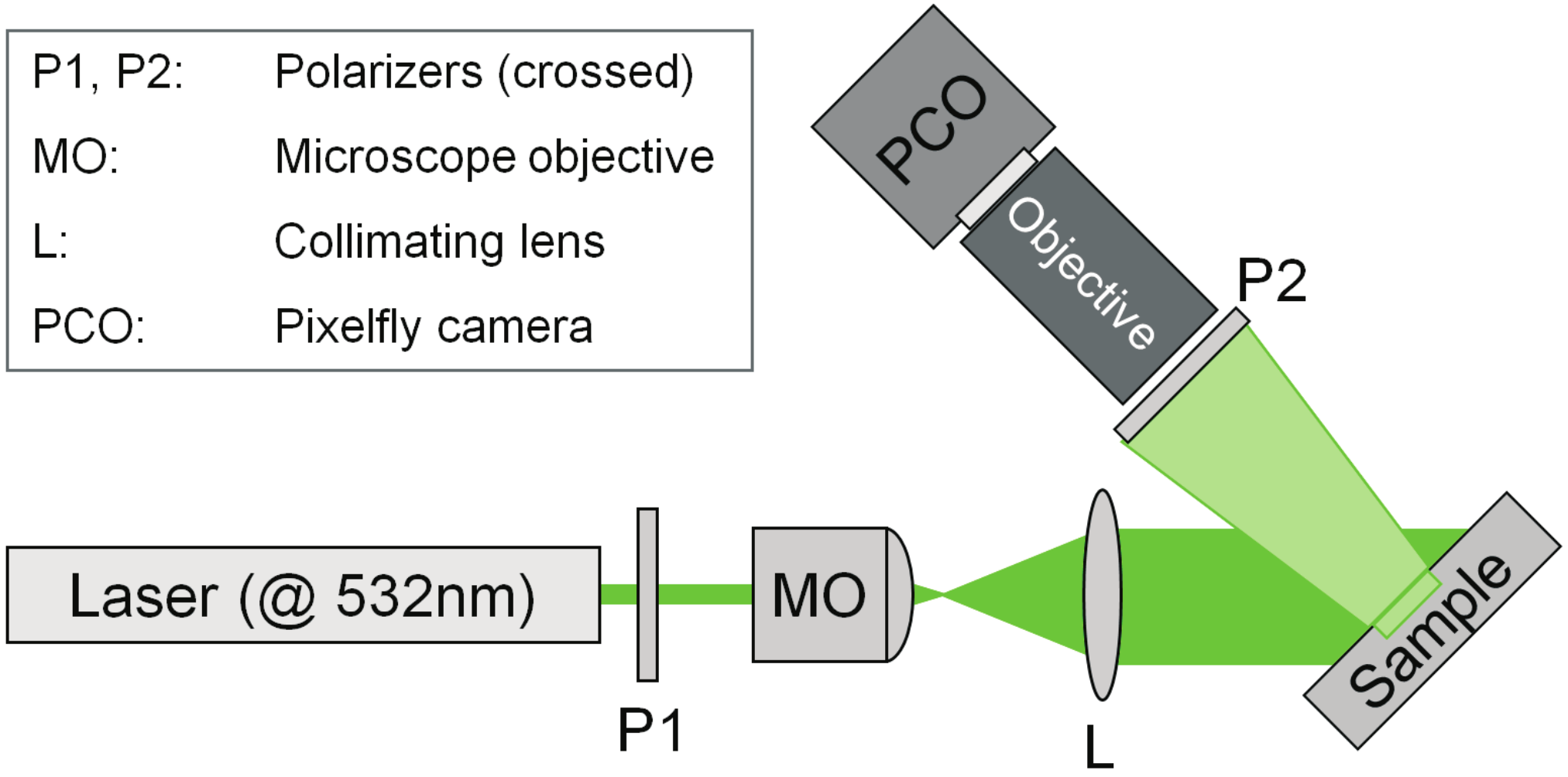}
\caption{Experimental setup. A flat sample is illuminated with a polarized expanded laser beam. The diffuse reflected light is detected in the cross polarization channel by imaging the surface of the sample with a digital camera. The size of individual speckle spots in the image plane can be adjusted by changing the aperture of the camera objective.}
\label{FigSetup}
}
\end{figure}

Consider coherent laser light scattered by a random sample in a typical light scattering experiment in reflection (see Fig.\ \ref{FigSetup}). Usually the cross polarized channel is analyzed in order to suppress specular and low order scattering reflection. A digital camera provides us with $N$ values of integrated (over the area of a pixel) intensity $I_{\alpha}$, $\alpha = 1, \ldots N$, corresponding to $N$ distinct pixels. $N$ could be a (typically, large) total number of available pixels, but it could also be just a small subset of pixels carrying information about only a part of the optical field, scattered by a small part of the sample (see Fig.\ \ref{FigGrid}). Our purpose is to characterize the statistical properties of the optical field based on this information. Note that quite generally the statistical properties of the fully developed speckle patterns considered here are the same for measurements in the image plane or in the far-field \cite{goodman}.  Therefore our results should find applications in many research areas where speckle based techniques are employed.

\subsection{Negative exponential distribution of integrated intensities}

We start by considering a speckle pattern resulting from scattering of light in a solid sample, where the positions of scattering centers are fixed.
It is known that under quite general conditions, the intensity $I$ of light scattered from such a random medium follows the negative exponential distribution: $P(I) = (1/\langle I \rangle) \exp(-I/\langle I \rangle)$ \cite{goodman}.
In the experiment we only have access to
\begin{eqnarray}
I_{\alpha} = \int\limits_{\mathrm{pixel}~\alpha} d^2 \mathbf{r}\, I(\mathbf{r}).
\label{intint}
\end{eqnarray}
which is the intensity integrated over an area $a^2$ of a pixel of the camera. Here $\alpha = 1, \ldots, N$ indexes different pixels of the camera.
As the first and the simplest example, we assume that $I_{\alpha}$ follows the negative exponential distribution as well and, in addition, that integrated intensities corresponding to different pixels are uncorrelated. The average value of intensity $I$ and the variance of its fluctuations can be estimated from the data $\left\{ I_{\alpha} \right\}$ as
\begin{eqnarray}
i &=& \frac{1}{N} \sum\limits_{\alpha = 1}^N I_{\alpha},
\label{int}
\\
c &=& \frac{1}{N-1} \sum\limits_{\alpha = 1}^N (I_{\alpha} - i)^2.
\label{var}
\end{eqnarray}
These statistical estimators are unbiased, i.e. the mean values of $i$ and $c$ found by averaging over an ensemble of realizations are equal to the actual values of the average intensity and of its variance, respectively:
\begin{eqnarray}
\langle i \rangle &=& \langle I \rangle,
\label{aint}
\\
\langle c \rangle &=& \langle (I - \langle I \rangle)^2 \rangle =
\langle I \rangle^2.
\label{avar}
\end{eqnarray}
Here the angular brackets $\langle \ldots \rangle$ denote ensemble averaging. The normalized variance $\langle c \rangle/\langle I \rangle^2$ is equal to one in this case. The \emph{speckle contrast} $K$ can be defined as
\begin{equation}
K = \sqrt{\langle c \rangle}/\langle I \rangle.
\end{equation}

The values of $i$ and $c$ obtained in a series of measurements fluctuate around their means (\ref{aint}) and (\ref{avar}). The variances of these fluctuations are
\begin{eqnarray}
\frac{\sigma_i^2}{\langle i \rangle^2} &=&
\frac{\langle i^2 \rangle - \langle i \rangle^2}{\langle i \rangle^2} = \frac{1}{N},
\label{varint}
\\
\frac{\sigma_c^2}{\langle c \rangle^2} &=&
\frac{\langle c^2 \rangle - \langle c \rangle^2}{\langle c
\rangle^2} = \frac{8}{N} \times \frac{N - \frac34}{N-1}
\simeq \frac{8}{N},
\label{varvar}
\end{eqnarray}
where in the last line, the result after the ``$\simeq$'' sign was obtained by taking the limit $N \rightarrow \infty$.

\begin{figure}[t]
\centering{
\includegraphics[width=8cm,angle=90]{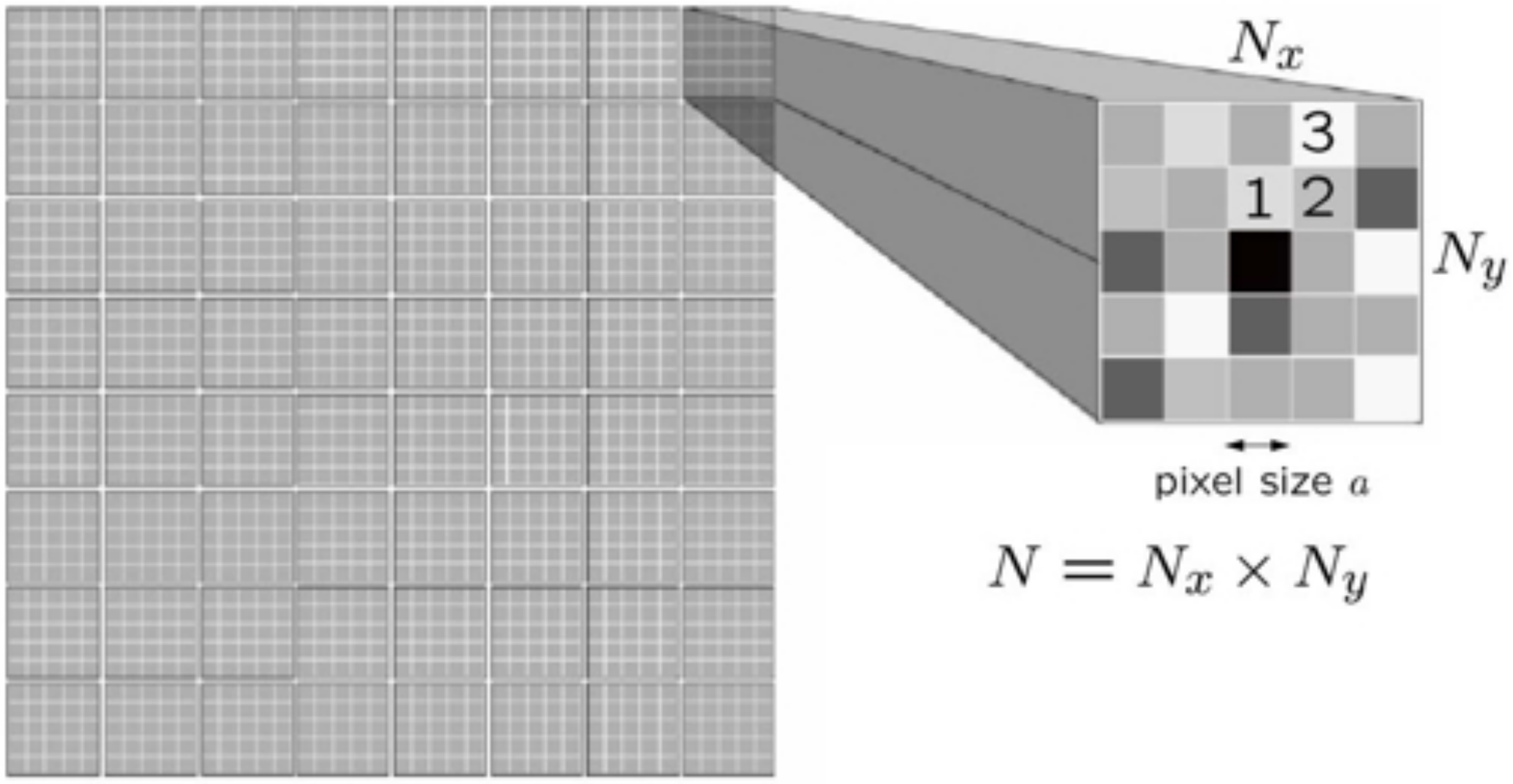}
\caption{Schematic representation of the matrix of pixels of a digital camera. The full matrix is divided in a set of meta-pixels $N_x \times N_y = N$ pixels each. Spatially-varying statistical properties of the speckle pattern imaged by the camera (i.e., the variance of intensities $c$) are estimated by averaging over all pixels within the same meta-pixel. In our calculation (Sec.\ \ref{SecCorr}), we are taking into account correlations between intensities at neighboring pixels in all directions. A given pixel (pixel 1 in the figure) have 8 neighbors: 4 neighbors of type 2 and 4 neighbors of type 3.}
\label{FigGrid}
}
\end{figure}

To which extent are Eqs.\ (\ref{aint}--\ref{varvar}) relevant to realistic experiments? For a standard square matrix of adjacent pixels (see Fig.\ \ref{FigGrid}), two situations are possible depending on the relation between the speckle size (i.e. the correlation range $b$ of the scattered intensity) and the pixel size $a$. For small speckles ($b \ll a$), different pixels are uncorrelated, but the statistical distribution of integrated intensity $I_{\alpha}$ differs from the negative exponential one. In the opposite case of large speckles ($b \gtrsim a$) the statistical distribution of $I_{\alpha}$ approaches the negative exponential one, but intensities corresponding to neighboring pixels become correlated. The situation described in the present section --- uncorrelated pixels with negative exponential distribution of $I_{\alpha}$ --- can be reached by working with large speckles ($b \gtrsim a$) but using only a subset of pixels (i.e., only pixels distanced by 1, 2 or even more pixels) in the analysis.
However, using only a subset of all available pixels means that the total number of pixels is reduced. It is therefore desirable to extend Eqs.\ (\ref{aint}--\ref{varvar}) to a more realistic model for the distribution function of $I_{\alpha}$.

\begin{figure}[t]
\centering{
\includegraphics[width=13cm,angle=0]{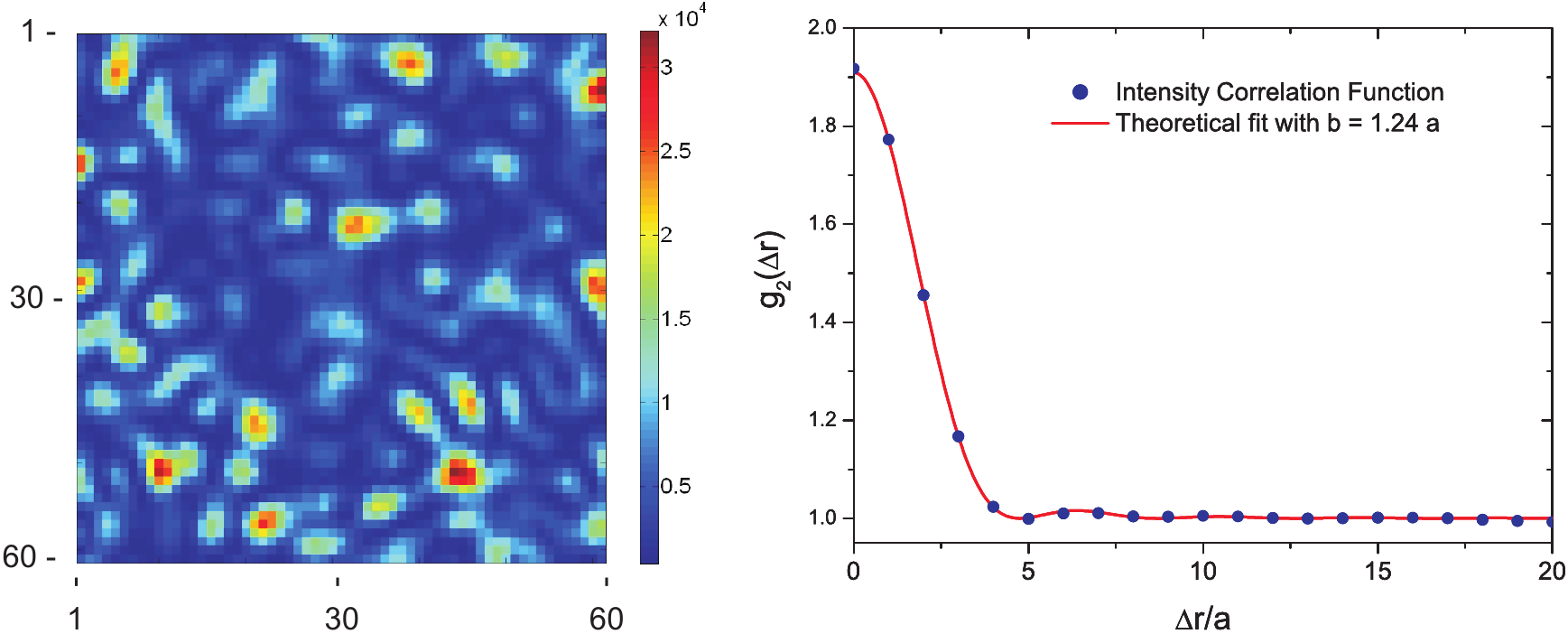}
\caption{Left: False-color image of light intensities in a speckle pattern ($60 \times 60$ pixels) obtained by using the smallest available aperture setting ($f/\#$=32). Intensity scale from $0$ to $10^4$ in arbitrary units. Right: Intensity correlation function obtained by fast Fourier transformation (FFT) of the raw speckle image (full frame $640 \times 480$ pixels). The data is quantitatively described by Eq.\ (\ref{intcorrcirc}) with $b = 1.24 a$ and $\mu=1.09$.}
\label{FigSpeckleSizeFFT}
}
\end{figure}

\subsection{Gamma distribution of integrated intensities}
\label{secgamma}

\begin{figure}[t]
\centering{
\includegraphics[width=10cm,angle=0]{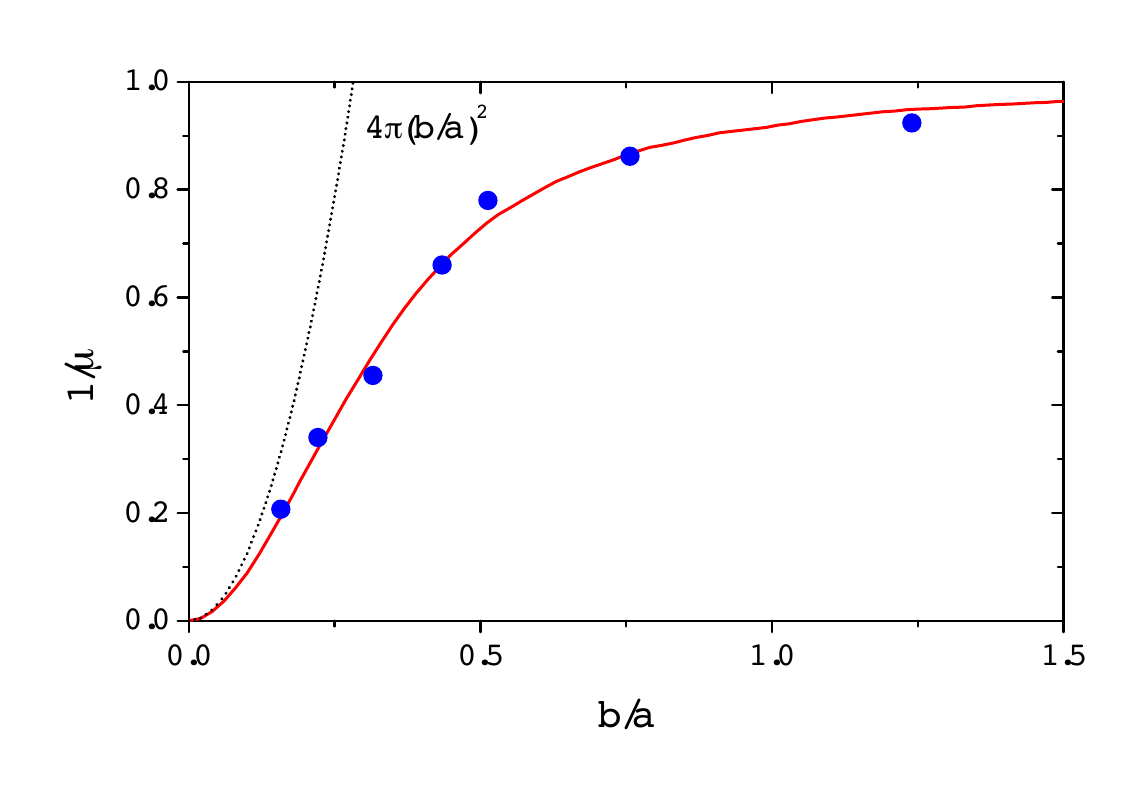}
\caption{Speckle parameter $\mu$ as a function of the speckle size $b$ divided by the size of the camera pixel $a$. Symbols: experimental results for the case of scattering from a solid sample (Teflon). Solid line: Eq.\ (\ref{mu}). Dashed line: the approximate result in the small-speckle limit $b \ll a$.}
\label{FigSpeckleSize}
}
\end{figure}

For small speckles ($b \ll a$) the integrated intensities $I_{\alpha}$ corresponding to different pixels can be considered independent, whereas the distribution of each $I_{\alpha}$, according to Goodman \cite{goodman},  can be modeled with the so-called gamma distribution :
\begin{eqnarray}
P(I_{\alpha}) = \frac{1}{\Gamma(\mu)} \left( \frac{\mu}{\langle I \rangle} \right)^{\mu}
I_{\alpha}^{\mu-1} \exp\left(-\frac{\mu I_{\alpha}}{\langle I \rangle} \right).
\label{gamma}
\end{eqnarray}
Here the parameter $\mu \geq 1$ depends on the pixel shape and size, as well as on the spatial correlation function $g_2(\Delta\mathbf{r})$ of intensity $I(\mathbf{r})$:
\begin{eqnarray}
\frac{1}{\mu} &=& \frac{1}{a^4}
\int\limits_{\mathrm{pixel}} d^2 \mathbf{r}
\int\limits_{\mathrm{pixel}} d^2 \mathbf{r}'
\left[ g_2 \left(\mathbf{r} - \mathbf{r}' \right) -1 \right],
\label{mu}
\end{eqnarray}
where both integrations run over the area of the same pixel.
If the sample is illuminated by a Gaussian beam, the intensity correlation function of the far-field speckle pattern is
\begin{eqnarray}
g_2(\Delta\mathbf{r}) = \frac{\langle I(\mathbf{r})  I(\mathbf{r}+\Delta\mathbf{r})\rangle}{
\langle I \rangle^2}
= 1 +
\exp\left(-\frac{\Delta \mathbf{r}^2}{b^2} \right),
\label{intcorrgauss}
\end{eqnarray}
whereas for a plane wave passed through a circular aperture,
\begin{eqnarray}
g_2(\Delta\mathbf{r})
= 1 + \left[ \frac{2 J_1(\Delta r/b)}{(\Delta r/b)} \right]^2.
\label{intcorrcirc}
\end{eqnarray}
In our experiments, the correlation function of intensity in the speckle pattern is controlled by the (circular) aperture of the camera objective, so that Eq.\ (\ref{intcorrcirc}) is more appropriate for us. In both cases $b$ quantifies the extent of spatial correlations of $I(\mathbf{r})$ (i.e., the size of a single speckle spot). An analytic expression for $\mu$ can only be obtained for very small speckles $b \ll a$:  $\mu \simeq (a/b)^2/(4\pi)$. In Fig.\ \ref{FigSpeckleSize} we compare the theoretical Eq.\ (\ref{mu}) with experimental results and find an excellent agreement.

The expectations and the variances of $i$ and $c$ defined by Eqs.\ (\ref{int}) and (\ref{var}) can be calculated for the gamma distribution using the known expression for its statistical moments:
\begin{eqnarray}
\langle I_{\alpha}^n \rangle = \frac{\Gamma(\mu + n)}{\Gamma(\mu)} \left( \frac{\langle I \rangle}{\mu} \right)^n.
\label{moments}
\end{eqnarray}
We obtain
\begin{eqnarray}
&&\langle i \rangle = \langle I \rangle,
\;\;\;\;\;
\frac{\sigma_i^2}{\langle i \rangle^2} = \frac{1}{\mu N},
\label{gammai}
\\
&&\langle c \rangle = \frac{\langle I \rangle^2}{\mu},
\label{gammac}
\\
&&\frac{\sigma_c^2}{\langle c \rangle^2} =  \frac{2}{N-1}
\left[1 + \frac{3}{\mu} \left(1- \frac{1}{N} \right) \right]
\simeq \frac{2}{N}
\left(1 + \frac{3}{\mu} \right),
\label{gammasigmax}
\end{eqnarray}
where, again, the last result of the second line is obtained in the limit of $N \rightarrow \infty$. As expected, these results coincide with Eqs.\ (\ref{aint}--\ref{varvar}) for $\mu = 1$, when Eq.\ (\ref{gamma}) becomes identical to the negative exponential distribution. When $\mu > 1$, the variance of intensities $c$ of the speckle pattern is reduced due to the effective spatial averaging over the area of a pixel that becomes comparable to the typical size of a speckle spot $b$ [see Eq.\ (\ref{gammac})]. At the same time, fluctuations of $c$ from one meta-pixel to another are suppressed by a factor $(1 + 3/\mu)/4 < 1$ [see Eq.\ (\ref{gammasigmax})].

Figure \ref{FigSpeckleSizeFFT} (left) shows a speckle image of light scattered from solid Teflon  taken for the case of large speckles that can be resolved in space ($f \mathord{\left/
{\vphantom {f \# }} \right.
\kern-\nulldelimiterspace} \# = 32$). Be $\tilde I( \mathbf{q})$ the Fourier transform of the image $I( \mathbf{r})$. From the inverse Fourier transform of the power spectrum $|I( \mathbf{q})|^2$ we directly obtain the non-normalized intensity correlation function $g_2(\Delta r)$ \cite{goodman} [see Fig.\ \ref{FigSpeckleSizeFFT} (right)]. In our example, for speckles of finite size $b/a=1.24$, the intercept of the correlation function is given by $1+1/\mu=1.92$.

\subsection{Role of correlations between neighboring pixels}
\label{SecCorr}

The results obtained in the previous subsection can be extended to include correlations between neighboring pixels. In a matrix of adjacent square pixels (Fig.\ \ref{FigGrid})
of side $a$ larger than the correlation range $b$ of the speckle field, it appears reasonable to take into account only correlations between integrated intensities $I_{\alpha}$ corresponding to neighboring pixels and to neglect correlations between intensities of more distant pixels. Note that each pixel has 8 neighboring pixels of two different types (pixels 2 and 3 in Fig.\ \ref{FigGrid}). The impact of correlations on the average values and variances of $i$ and $c$ is described by two additional parameters
\begin{eqnarray}
\frac{1}{\mu_2} &=& \frac{1}{a^4}
\int\limits_{\mathrm{pixel}~1} d^2 \mathbf{r}
\int\limits_{\mathrm{pixel}~2} d^2 \mathbf{r}'
\left[ g_2 \left(\mathbf{r} - \mathbf{r}'\right) -1 \right],
\label{mu2}
\\
\frac{1}{\mu_3} &=& \frac{1}{a^4}
\int\limits_{\mathrm{pixel}~1} d^2 \mathbf{r}
\int\limits_{\mathrm{pixel}~3} d^2 \mathbf{r}'
\left[ g_2 \left(\mathbf{r} - \mathbf{r}'\right) -1 \right],
\label{mu3}
\end{eqnarray}
where, in contrast to Eq.\ (\ref{mu}), the two integrations in each equation run over two different pixels: the neighboring pixels 1 and 2 situated side by side in Eq.\ (\ref{mu2}) or the pixels 1 and 3 situated next to each other on a diagonal of a square matrix of pixels in Eq.\ (\ref{mu3}), see Fig.\ \ref{FigGrid}.

With this definitions in hand, we can readily calculate the average value of $i$, $\langle i \rangle = \langle I \rangle$, and its variance:
\begin{eqnarray}
\frac{\sigma_i^2}{\langle i \rangle^2} = \frac{1}{N}
\left[ \frac{1}{\mu} +
\frac{4}{\mu_2} \left(1 - \frac{1}{\sqrt{N}} \right) +
\frac{4}{\mu_3} \left(1 - \frac{2}{\sqrt{N}}
+ \frac{1}{N} \right)  \right].
\label{varcorr}
\end{eqnarray}
The average value of $c$ can also be readily found:
\begin{eqnarray}
\langle c \rangle = \langle I \rangle^2
\left[ \frac{1}{\mu} -
\frac{4}{\mu_2 \sqrt{N}(\sqrt{N} + 1)} -
\frac{4 (\sqrt{N} - 1)}{\mu_3 N (\sqrt{N} + 1)}
\right].
\label{avccorr}
\end{eqnarray}
For $N  \to \infty$, Eq.\ (\ref{avccorr}) reduces to Eq.\ (\ref{gammac}).

To obtain these results we gave special treatment to pixels situated at the boundaries or in the corners of the square matrix. These pixels have less neighbors, which is important when Eqs.\ (\ref{varcorr}) and (\ref{avccorr}) are used for small numbers of pixels $N \sim 10$.
 An interesting consequence of Eq.\ (\ref{avccorr}) is that the expectation value of $c$ becomes $N$-dependent. Because we saw in the previous sections that when different pixels are considered independent, $\langle c \rangle$ does not depend on $N$ [cf. Eqs.\ (\ref{avar}) and (\ref{gammac})], we conclude that the dependence of $\langle c \rangle$ on $N$ is a signature of correlation between  signals corresponding to different pixels. The estimator (\ref{var}) acquires a bias. Interestingly, the spatial range of correlations $b$ or, more precisely, the ratio $b/a$ can be estimated by observing the $N$-dependence of $\langle c \rangle$.

The variance of $c$, $\sigma_c^2$, can be calculated along the same lines as $\langle c \rangle$. The calculation, however, is much more involved and rapidly leads to cumbersome equations. Instead of dealing with this lengthy and, anyway, approximate analytic calculation, we estimate $\sigma_c^2$ by a numerical simulation.
We use a computer to generate a set of 4096$\times$4096 speckle patterns with correlation lengths $b = 2$, 4, and 8, respectively. These speckle patterns are then used to generate integrated speckle patterns with pixel sizes $a$ varying from 2 to 64. The integrated speckle patterns model outputs of a typical CCD or CMOS camera. Finally, we numerically compute the dependencies of $\langle c \rangle$ and $\sigma_c^2$ on the number $N$ of pixels, for fixed ratios $b/a$. For $b/a \ll 1$ the parameter $\mu$ characterizing integrated speckle patterns [Eq.\ (\ref{mu})] is a function of $b/a$ only. Hence, by inverting this function we can express $b/a$ and hence $\sigma_c^2/\langle c \rangle^2$ as a function of $\mu$ instead of $b/a$. We know that Eq.\ (\ref{gammasigmax}) is valid in the limit of $\mu \rightarrow \infty$ (or, equivalently, $b/a \rightarrow 0$) which suggests that $\sigma_c^2/\langle c \rangle^2$ can be expressed as a series in $1/\mu$ of which the terms of order $(1/\mu)^0$ and $(1/\mu)^1$ are already known from Eq.\ (\ref{gammasigmax}). In the limit of large $N$, the results of our simulations can be fit by adding a term quadratic in $1/\mu$:
\begin{eqnarray}
\frac{\sigma_c^2}{\langle c \rangle^2} &\simeq&  \frac{2}{N}
\left[1 + \frac{3}{\mu} + \frac{12}{\mu^2} \right] =
\frac{H(\mu)}{N},
\label{gammasigmax2}
\end{eqnarray}
where we introduced
$H(\mu) \simeq 2(1 + 3/{\mu} + 12/{\mu^2})$. This equation appears to describe our results quite well for $0 < 1/\mu < 0.8$ (see Sec.\ \ref{exp}). For uncorrelated pixels $H(\mu) \simeq 2(1 + 3/{\mu})$ from Eq. (\ref{gammasigmax}).

\begin{figure}[t]
\centering{
\includegraphics[width=13cm,angle=0]{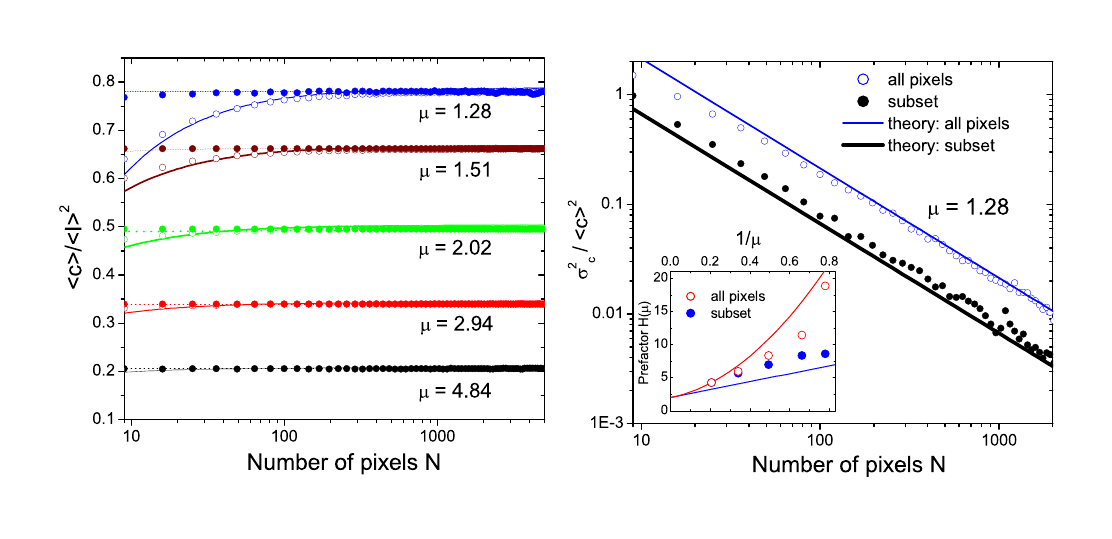}
\caption{Average variance of the intensity fluctuations $\langle c \rangle$ and its noise $\sigma^2_c$ as functions of the number of pixels $N$. Speckle pattern is recorded in the image plane for light reflected from a solid piece of Teflon. Exposure time is 1 ms, transport mean free path in Teflon $l^*\simeq 0.25$ mm, camera pixel size $ a = 9.9$ $\mu$m, magnification one. Left panel: analysis using a subset of pixels (all neighboring pixels omitted, full symbols) leads to a constant value of $\langle c \rangle$ [Eq.\ (\ref{gammac}), dotted lines]. Analysis using all pixels (open symbols) is compared to the prediction of Eq.\ (\ref{avccorr}) (solid lines). Right panel: dotted and solid lines show predictions of Eqs.\ (\ref{varvar}) and (\ref{gammasigmax2}), respectively. The inset shows a comparison of the experimental values (symbols) and theoretical predictions for $H(\mu) = N \sigma_c^2/\langle c \rangle^2$: Eq.\ (\ref{gammasigmax}) (dotted line) and Eq.\ (\ref{gammasigmax2}) (solid line).}
\label{FigTeflon}
}
\end{figure}

\subsection{Properties of time-integrated speckle patterns}
\label{timeint}

If the scattering medium is not stationary like, e.g., a liquid suspension of dielectric particles, the intensity $I$ of scattered light fluctuates in time. In such an experiment  the intensity $I$ of scattered light changes not only as a function of position $\mathbf{r}$, but also as a function of time $t$: $I = I(\mathbf{r}, t)$.
Fluctuations of $I$ in time can be characterized by its autocorrelation function,
\begin{eqnarray}
g_2(\tau) &=& \frac{\langle I(t) I(t+\tau) \rangle}{\langle I \rangle^2},
\label{g2}
\end{eqnarray}
where we omitted the position $\mathbf{r}$ from the arguments of $I$, for clarity. The analysis presented in the previous section applies to this situation too, provided that the sampling time $T$ (i.e. the time during which the signal is accumulated to determine $I_{\alpha}$) is much shorter than the characteristic correlation time $\tau_{\mathrm{c}}$ of variations of $I$.
In the opposite case, i.e. when $T \gtrsim \tau_{\mathrm{c}}$, the definition of $I_{\alpha}$ should include not only the spatial integration over the pixel area, but the temporal integration over the sampling time as well.

Under the assumption of small pixel size $a \ll b$ and statistically independent $I_{\alpha}$ following the negative exponential distribution, the effect of finite sampling time can be described by a parameter $\nu$
\cite{schatzel:noise90,durian:svs2005}:
\begin{equation}
\frac{1}{\nu} = \frac{2}{T} \int_0^T \left[ g_2(\tau) - 1 \right] \left ( 1 - \tau / T
\right ) d\tau.
\label{nu}
\end{equation}
The integrated intensity $I_{\alpha}$ follows the gamma distribution (\ref{gamma}) with $\nu$ substituted for $\mu$ \cite{goodman}. The average value $\langle c \rangle$ and its normalized variance are given by Eqs.\ (\ref{gammac}) and (\ref{gammasigmax}), respectively.

For pixel size $a$ that becomes comparable to the correlation length $b$ of the speckle pattern, we can take into account correlations between integrated intensities corresponding to neighboring pixels following the approach of Sec.\ \ref{SecCorr}. The outcome of our analysis is quite simple: all results of Sec.\ \ref{SecCorr} --- and, in particular, Eqs.\ (\ref{varcorr}), (\ref{avccorr}) and (\ref{gammasigmax2}) --- still apply but with $\mu$, $\mu_2$ and $\mu_3$ replaced by $\mu \times \nu$, $\mu_2 \times \nu$ and $\mu_3 \times \nu$, respectively. Here $\mu$, $\mu_2$ and $\mu_3$ describe the impact of spatial correlations, whereas $\nu$ accounts for time integration. The effects of spatial and time correlations thus fully decouple. This is due to the fact that the spatio-temporal correlation function $g_2(\Delta \mathbf{r}, \tau)-1 = \langle I(\mathbf{r}, t) I(\mathbf{r}+\Delta \mathbf{r}, t+\tau)\rangle/\langle I \rangle^2-1$ decouples, in its turn, into a product of position- and time-dependent parts.

\subsection{Comparison with experiment}
\label{exp}

In our experiment, linearly polarized light from a solid state laser (Verdi V5 from Coherent, wavelength $\lambda = 532$~nm) is expanded and collimated to a beam of several centimeters in waist to create an approximately homogeneous illumination spot on the sample surface (see Fig.\ \ref{FigSetup}). Liquid samples are contained in large glass containers whereas solid samples are measured in air. The diffuse reflected light is monitored in the image plane in the crossed polarization channel with a CCD camera PCO Pixelfly (640$\times$480 pixels of
9.9$\times$9.9~\textmu m$^2$ size, 12 bit) at magnification one. The estimated depth-of-focus of the imaging system is
close to 0.1 mm. The actual size of speckle spots on the CCD chip can be adjusted by varying the aperture of the camera objective. For the simple configuration of a single lens placed at a distance $l$ from the screen and a circular aperture with radius $q$ the speckle correlation length is \cite{Yoshimara1986} $b = {{2l} \mathord{\left/
 {\vphantom {{2l} q}} \right.
 \kern-\nulldelimiterspace} q} = \left[ {{{4l} \mathord{\left/
 {\vphantom {{4l} {kf}}} \right.
 \kern-\nulldelimiterspace} {kf}}} \right] \times {f \mathord{\left/
 {\vphantom {f \# }} \right.
 \kern-\nulldelimiterspace} \# } \label{specklesizeYoshi}$.

Our camera objective has an equivalent focal length of $f=50$ mm and is placed at a distance $l \sim 100$ mm from the CCD sensor. The $f$-number of the objective can be varied from $f \mathord{\left/
 {\vphantom {f \# }} \right.
 \kern-\nulldelimiterspace} \# = 2.8$ to $32$. The accessible values of $b$ are thus comparable or smaller than the pixel size $b \le  a$.
 The camera exposure time is adjustable with the shortest time used of 0.1 ms. The maximum speed of data acquisition at full resolution is 50 frames per second.
The recorded images are corrected for slight spatial variations of the average intensity and, subsequently, the statistical properties are calculated from the recorded intensity values. We have carefully checked that this correction procedure does not alter the experimental results. Throughout this article we compare two different types of experiments with theory. In an idealized experiment we analyze only a \emph{subset} of pixels where neighboring pixels are omitted.  The procedure effectively eliminates the effect of spatial correlations between the pixels for all situations considered here. The second set of data is obtained by analyzing all available pixels, a situation typically encountered in actual applications.

We first analyze the static speckle pattern of light reflected from a solid piece of Teflon (thickness $\approx $1 cm) with a typical photon transport mean free path $l^*\simeq 0.25$ mm. In Fig.\ \ref{FigTeflon},  we compare our experimental results for $\langle c \rangle$ (left panel) and its normalized variance $\sigma_c^2/\langle c \rangle^2$ (right panel) with the predictions of the theoretical model developed above. The parameter $\mu$ is determined from a best fit to the data via the relation $\mu  = {{\left\langle I \right\rangle ^2 } \mathord{\left/
 {\vphantom {{\left\langle I \right\rangle ^2 } {\left\langle c \right\rangle }}} \right.
 \kern-\nulldelimiterspace} {\left\langle c \right\rangle }}$ for $N \to \infty$. The values obtained for $\mu$ are found in excellent agreement with the theoretical predictions as shown in Fig.\ \ref{FigSpeckleSize}. Here the parameter $b$ for all settings has been obtained from the relation $b \propto f \mathord{\left/
 {\vphantom {f \# }} \right.
 \kern-\nulldelimiterspace} \# $ using the known value of $b=1.24$ for $f \mathord{\left/
 {\vphantom {f \# }} \right.
 \kern-\nulldelimiterspace} \# =32$ (Fig.\ \ref{FigSpeckleSizeFFT}). For consistency, in the following discussion for dynamic media, we use the fitted values of $\mu$ for each setting.
  
 For the static samples, good agreement with theory is found for the $N-$ dependence of the mean and the variance of $\langle c \rangle$. In particular, we clearly see that $\langle c \rangle$ is indeed independent of $N$ for a subset of uncorrelated pixels, whereas it acquires an $N$-dependence for correlated pixels. The expected $1/N$ dependence of the variance of $c$ is observed for both uncorrelated and correlated pixels, though with different prefactors. As follows from Eq.\ (\ref{gammasigmax2}), the prefactor $H(\mu)$ depends on the degree of correlation between neighboring pixels that we quantify by the parameter $\mu$. It is shown in the inset of Fig.\ \ref{FigTeflon}. The values of $H$ following from the experiment are close to those expected theoretically. Small but visible discrepancies between data and theory in Fig.\ \ref{FigTeflon} are most probably due to the approximate nature of our theoretical model: even for uncorrelated pixels the statistical distribution of intensity does not follow the Gamma distribution exactly \cite{goodman} and weak correlations certainly exist not only between neighboring but between distant pixels as well.

\section{Noise in photon correlation spectroscopy}

\subsection{Fluctuations of the intensity correlation function}
\label{secicf}

We now turn our attention to imaging and spectroscopy with speckle correlations. Instead of integrating the intensity of light scattered from a turbid sample with mobile scattering centers over a time exceeding the correlation time $\tau_c$ of intensity fluctuations (as we did in Sec.\ \ref{timeint}), detailed information on temporal fluctuations of intensity can be obtained by choosing $T \ll \tau_c$ and correlating speckle images separated by a time lag $\tau$ larger than $T$ . When intensities $I_{\alpha}(t)$ are measured for $N$ pixels, the degree of intensity correlation can be estimated as
\begin{eqnarray}
c(\tau) = \frac{1}{N-1} \sum\limits_{\alpha = 1}^N
\left[ I_{\alpha}(t) - i(t) \right]
\left[ I_{\alpha}(t+\tau) - i(t+\tau) \right],
\label{ctime}
\end{eqnarray}
where
\begin{eqnarray}
i(t) = \frac{1}{N} \sum\limits_{\alpha = 1}^N I_{\alpha}(t)
\label{inttime}
\end{eqnarray}
is an unbiased estimator of average intensity $\langle I \rangle$; it is equivalent to $i$ of Sec.\ \ref{static}. However, it will be important for us to keep track of the difference between $i(t)$ and $i(t+\tau)$ in the definition (\ref{ctime}) of $c(\tau)$.

The average and the variance of $c(\tau)$ for independent pixels with $I_{\alpha}(t)$ distributed according to the negative exponential distribution are:
\begin{eqnarray}
\langle c(\tau) \rangle &=& \langle I \rangle^2
\left[ g_2(\tau) - 1 \right],
\label{avctime}
\\
\frac{\sigma_{c(\tau)}^2}{\langle c(\tau) \rangle^2} &=&
\frac{g_2(\tau)^2 \left( 3-\frac{2}{N} \right)
- 2 \left[ g_2(\tau) - \frac{1}{N} \right]}{(N-1)
\left[ g_2(\tau)-1 \right]^2}
\simeq \frac{1}{N} \times \frac{g_2(\tau) \left[ 3 g_2(\tau) - 2 \right]}{\left[ g_2(\tau) - 1 \right]^2},
\label{varctime}
\end{eqnarray}
where the last expression is obtained in the limit of $N \rightarrow \infty$.

As we saw in the previous sections, it is important to account for deviations of the intensity distribution from the negative exponential one as well as for correlations between neighboring pixels in the image, if a comparison with experiment is attempted. Because the intensity correlation function $g_2(\Delta \mathbf{r}, \tau)-1 = \langle I(\mathbf{r}, t) I(\mathbf{r}+\Delta \mathbf{r}, t+\tau)\rangle/\langle I \rangle^2-1$ decouples into a product of position- and time-dependent parts, we readily obtain
\begin{eqnarray}
\langle c(\tau) \rangle &=& \langle I \rangle^2
\left[ \frac{1}{\mu} -
\frac{4}{\mu_2 \sqrt{N}(\sqrt{N} + 1)} -
\frac{4 (\sqrt{N} - 1)}{\mu_3 N (\sqrt{N} + 1)}
\right]
\left[ g_2(\tau) - 1 \right]
\label{avctime2}
\end{eqnarray}
for the average of $c(\tau)$. Here $\mu$, $\mu_2$ and $\mu_3$ account for spatial correlations of the instantaneous speckle pattern and are given by Eqs.\ (\ref{mu}), (\ref{mu2}) and (\ref{mu3}), respectively.

The calculation of the variance of fluctuations of $c(\tau)$ with account for correlations between neighboring  pixels appears to be much more involved. However, an approximate result can still be obtained by an \emph{ad hoc} combination of Eqs.\ (\ref{gammasigmax2}) and (\ref{varctime}):
\begin{eqnarray}
\frac{\sigma_{c(\tau)}^2}{\langle c(\tau) \rangle^2} &\simeq& \frac{H(\mu)}{8 N} \times \frac{g_2(\tau) \left[ 3 g_2(\tau) - 2 \right]}{\left[ g_2(\tau) - 1 \right]^2}.
\label{varctime2}
\end{eqnarray}

\begin{figure}[t]
\centering{
\includegraphics[width=13cm,angle=0]{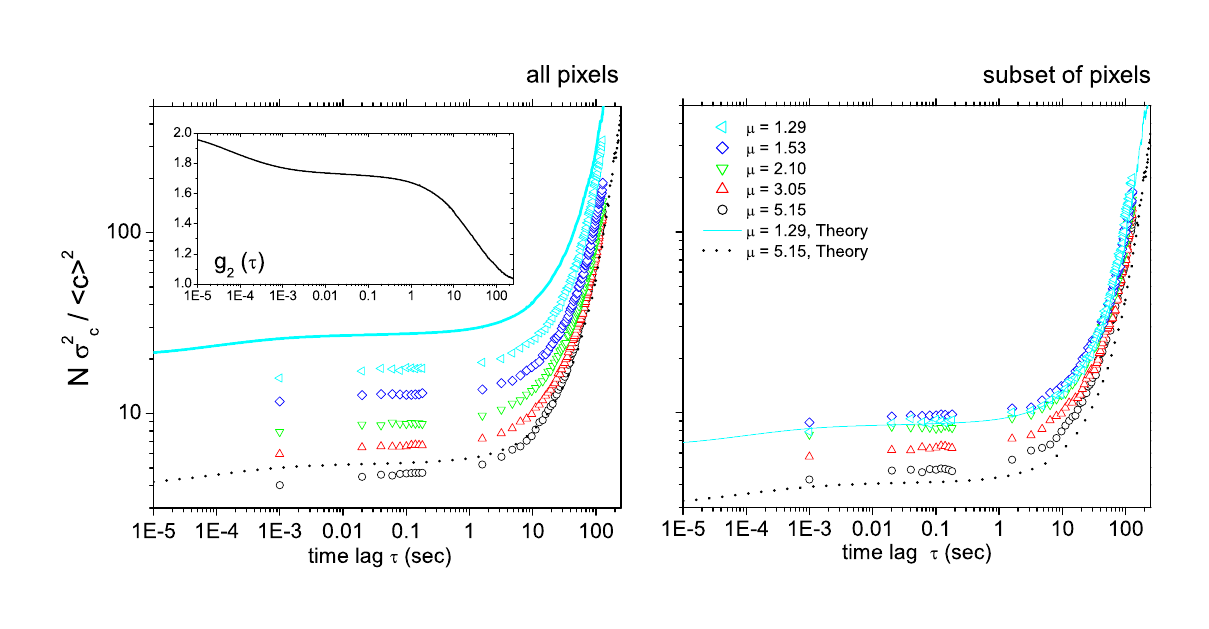}
\caption{Noise of the speckle correlation coefficient as a function of the time lag $\tau$ (symbols). Solid line shows prediction by Eqs.\ (\ref{varctime2}) (left) and (\ref{varctime}) (right). Inset of the left panel shows the intensity correlation function $g_2(\tau)$.}
\label{ICFDynamics}
}
\end{figure}

To compare Eqs.\ (\ref{avctime})--(\ref{varctime2}) with measurements we have carried out a time correlation experiment on a slowly relaxing sample.  We dispersed about 3.3$\%$ of polystyrene microspheres (diameter 710 nm) in an aqueous solution of cetylpyridinium chloride/sodium salicylate (100 mM
CPyCl/60 mM NaSal). The resulting surfactant solution strongly scatters light with the transport mean free path $l^* \simeq  73$ $\mu$m at 532 nm. A detailed characterization of the system is given in Ref.\ \cite{WLMPRL07}. It displays strongly viscoelastic properties with a  slow terminal relaxation. For our measurement  we keep the sample at room temperature $T \simeq 22^\circ$ C  which sets the relaxation time to several tens of seconds. We first determine the full autocorrelation function from a combined photon PCS and CCD-camera based experiment \cite{PineRSI2002}. This yields the temporal autocorrelation function $g_2(\tau)$ of scattered light (see the inset of Fig.\ \ref{ICFDynamics}). We then perform measurements of the noise of speckle correlation coefficient $c(\tau)$ for different sizes of the camera objective aperture corresponding to the values of $\mu$ ranging from 1.29 to 5.15, and as a function of the number of pixels $N$. In all cases we observe that the normalized variance of
$c(\tau)$ scales with $1/N$ as expected from the theory. In Fig.\ \ref{ICFDynamics} we show $N \sigma_{c}^2/\langle c \rangle^2$ as a function of the time lag $\tau$. The general rule is that the normalized variance of the correlation coefficient $c(\tau)$ increases with $\tau$.
Experiment and theory are in fairly good agreement if the analysis is performed on a subset of pixels (right panel of Fig.\ \ref{ICFDynamics}), especially at small $\mu$, when the statistics of intensity fluctuations is close to negative exponential. The agreement is worse when all pixels are analyzed (left panel of Fig.\ \ref{ICFDynamics}), even though the theory describes the general trend of our data quite well. This might either suggest that we were either unable to eliminate residual experimental errors or that additional factors, not taken into account in our theoretical model, enter into play when important correlations between signals corresponding to different pixels are present.

\subsection{Fluctuations of the intensity structure function}
\label{secisf}

Instead of the intensity correlation function $g_2(\tau)$, fluctuations of intensity can be characterized by a \textit{structure} function
\begin{eqnarray}
D(\tau) = \langle \left[ I(t) - I(t+\tau) \right]^2 \rangle
= \langle I \rangle^2 d(\tau),
\label{isf}
\end{eqnarray}
where for the case of Gaussian statistics
\begin{eqnarray}
d(\tau) = \frac{\langle \left[ I(t) - I(t+\tau) \right]^2 \rangle}{\langle I \rangle^2}
= 2 \left[g_2(0) - g_2(\tau) \right].
\label{isf}
\end{eqnarray}
The intensity structure function (ISF) is a more direct measure of the dynamic activity and has several advantages as compared to the intensity correlation function (ICF)
\cite{Schatzel83,Zakharov2010}. While both quantities are directly related in the limit of
perfect measurement statistics, the ISF is known to outperform the ICF in accuracy when the collection time is limited. Furthermore, the ISF is less sensitive to low frequency noise or drifts \cite{Schatzel83}. An unbiased estimator of $D(t)$ is
\begin{eqnarray}
s(\tau) = \frac{1}{N} \sum\limits_{\alpha = 1}^N
\left[ I_{\alpha}(t) - I_{\alpha}(t + \tau) \right]^2.
\label{isfest}
\end{eqnarray}

\begin{figure}[t]
\centering{
\includegraphics[width=13cm]{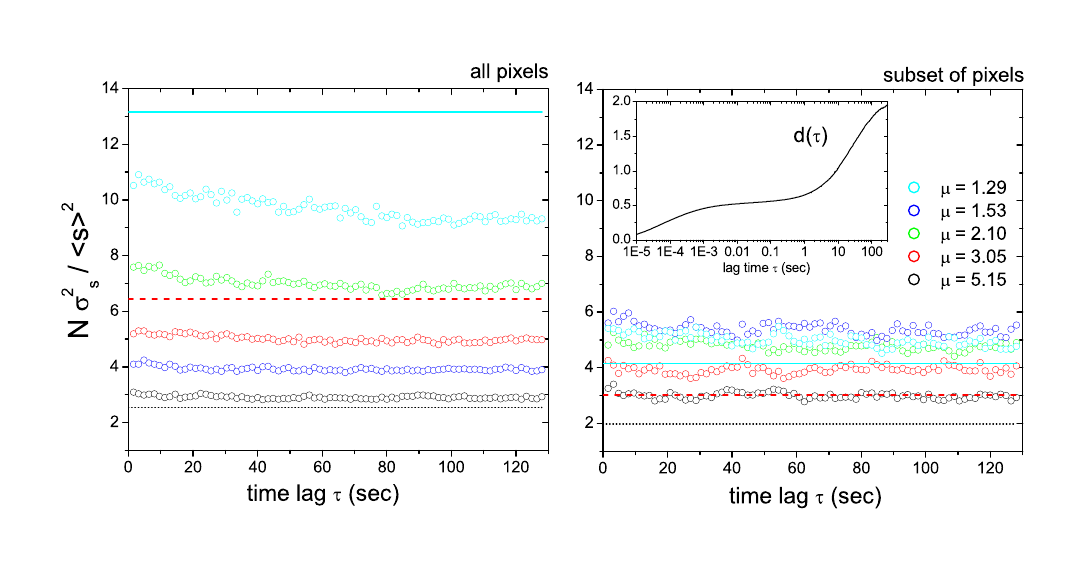}
\caption{Noise of the of the speckle structure coefficient as a function of time lag $\tau$ (symbols). As predicted by the theory, the noise is independent of $\tau$. Lines show predictions of Eqs.\ (\ref{dvar2}) (left) and (\ref{dvar}) (right). The inset of the right panel shows the intensity structure function $d(\tau)$.}
\label{ISFDynamics}
}
\end{figure}

For the negative exponential distribution of integrated intensities $I_{\alpha}(t)$, assuming no correlation between intensities at different pixels, we obtain
\begin{eqnarray}
\langle s(\tau) \rangle &=& D(\tau),
\label{dav}
\\
\frac{\sigma_{s(\tau)}^2}{\langle s(\tau) \rangle^2} &=& \frac{5}{N}.
\label{dvar}
\end{eqnarray}

Accounting for finite pixel size $a$ that can become comparable to the correlations length $b$ of the speckle pattern, we obtain
\begin{eqnarray}
\langle s(\tau) \rangle &=& \frac{D(\tau)}{\mu}.
\label{dav2}
\end{eqnarray}

Once again, the calculation of the variance of fluctuations of $s(\tau)$ taking into account correlations between neighboring  pixels appears to be much more involved. However, an approximate result can still be obtained by a (though \emph{ad-hoc}) combination of Eqs.\ (\ref{gammasigmax2}) and (\ref{dvar}):
\begin{equation}
\frac{\sigma_{s(\tau)}^2}{\langle s(\tau) \rangle^2} \simeq \frac{5}{{8N}}H(\mu).
\label{dvar2}
\end{equation}

A comparison between Eqs.\ (\ref{dvar}), (\ref{dvar2}) and experimental data is presented in Fig.\ \ref{ISFDynamics}. The theory correctly predicts that the noise of $s(\tau)$ scales with $1/N$ and that it is independent of $\tau$. However, theory and experiment do not show quantitative agreement: theoretical lines lie either higher (blue and red lines in the left panel of Fig.\ \ref{ISFDynamics}) or lower (all other lines in Fig.\ \ref{ISFDynamics}) than the data. The reasons behind this disagreement are the same as in the case of intensity correlation coefficient $c(\tau)$ in Sec.\ \ref{secicf}.

\section{Conclusions}

In the present paper we presented a study of noise in modern speckle correlation and imaging techniques. The noise originates from replacing the ensemble averaging (assumed in theoretical models) by averaging over a finite number $N$ of pixels of a digital camera. In particular, we studied the noise of the speckle intensity variance $c$, and of the intensity correlation and structure coefficients $c(\tau)$ and $s(\tau)$, respectively. The variances of all these quantities decrease as $1/N$ when the number of pixels is increased and depend in complex ways on the spatial (due to the finite pixel size) and temporal (due to the finite sampling time) correlations of intensity in the speckle pattern. For stationary speckle patterns, we obtained quantitative agreement between measurements and the theoretical model that we developed in this paper. For dynamic speckle patterns, theoretical predictions  reproduce general trends of our data but fail to provide a fully quantitative description. We believe that this is due to the approximate character of our theoretical model (neglecting correlations between distant pixels, gamma distribution of intensity at a single pixel) as well as to imperfections of our experiment (impossibility to achieve a perfectly uniform illumination of the sample, necessity to work with weak signals, etc.). Despite the absence of quantitative agreement between theory and experiment in the latter case, the results presented in the paper provide an important starting point for estimation of the noise level in such applications as the multi-speckle dynamic light scattering, time-resolved correlation spectroscopy, speckle visibility spectroscopy, near-field scattering, laser speckle imaging and echo speckle imaging.

\section*{Acknowledgements}

This work has been supported by the Swiss-French Germaine de Sta\"{e}l collaboration project No.~19126RL and the Swiss National Science Foundation projects 200020-
126772 and 117762. RC acknowledges financial support from the European Union (Marie Curie Intra-European Fellowship, Contract No. EIF-038772)


\begin{thebibliography}{99}


\bibitem{goodman}
J.W. Goodman, \emph{Speckle Phenomena in Optics} (Roberts \& Company, Englewood, Colorado, 2007).

\bibitem{Berne:DLS}
B.J. Berne and R. Pecora, \emph{Dynamic Light Scattering} (Wiley, New York, 1976).

\bibitem{Schatzel83}
K. Sch\"atzel, ``Noise in photon correlation and photon structure functions,'', J. Mod. Opt \textbf{30}, 155-166 (1983).

\bibitem{Kirsch96}
S. Kirsch, V. Frenz, W. Schartl, E. Bartsch, and H. Sillescu, ``Multispeckle autocorrelation spectroscopy and its application to the investigation of ultraslow dynamical processes,'' J. Chem. Phys. \textbf{104,} 1758-1761 (1996).

\bibitem{luca2000}
L. Cipelletti, S. Manley, R.C. Ball, and D.A. Weitz, ``Universal aging features in the restructuring of fractal colloidal gels,'' Phys. Rev. Lett. \textbf{84,} 2275-2278 (2000).

\bibitem{harden2000}
A. Knaebel, M. Bellour, J.P. Munch, V. Viasnoff, F. Lequeux, and J.L. Harden, ``Aging behavior of Laponite clay particle suspensions,'' Europhys. Lett. \textbf{52,} 73-79 (2000).

\bibitem{PineRSI2002}
V. Viasnoff, F. Lequeux, and D.J. Pine, ``Multispeckle diffusing-wave spectroscopy: a tool to study slow relaxation and time-dependent dynamics,'' Rev. Sci. Instrum. \textbf{73,}  2336-2344 (2002).

\bibitem{PavelAppop2006} P. Zakharov, S. Bhat, P. Schurtenberger,
F. Scheffold, ``Multiple scattering suppression in dynamic light scattering based on a digital camera detection scheme'',  \textbf{45,} 1756-1764 (2006)

\bibitem{CipTRC2003}
L. Cipelletti, H. Bissig, V. Trappe, P. Ballesta, and S. Mazoyer,
``Time-resolved correlation: a new tool for studying temporally heterogeneous dynamics,'' J. Phys. Cond. Mat. \textbf{15,} 257-262 (2003).

\bibitem{durian}
P.K. Dixon and D.J. Durian, ``Speckle visibility spectroscopy and variable granular fluidization,'' Phys. Rev. Lett. \textbf{90,} 184302 (2003).


\bibitem{durian:svs2005}
R. Bandyopadhyay, A.S. Gittings, S.S. Suh, P.K. Dixon, and D.J. Durian, ``Speckle-visibility spectroscopy: a tool to study time-varying dynamics,'' Rev. Sci. Instrum. \textbf{76,} 093110 (2005).


\bibitem{Cerbino2009}
R. Cerbino and A. Vailati, ``Near-field scattering techniques: novel instrumentation and results from time and spatially resolved investigations of soft matter systems,'' Curr. Opin. Colloid Interface Sci. \textbf{14,} 416-425 (2009).

\bibitem{briers96:lasca}
J.D. Briers and S. Webster, ``Laser speckle contrast analysis (LASCA): a nonscanning, full-field technique for monitoring capillary blood flow,''
J. Biomed. Opt. \textbf{1,} 174-179 (1996).

\bibitem{Pavel2009OE} P. Zakharov, A.C. Volker, M.T. Wyss, F. Haiss, N. Calcinaghi, C. Zunzunegui, A. Buck, F. Scheffold and B. Weber, ``Dynamic laser speckle imaging of cerebral blood flow,'' Optics Express  \textbf{16,} 13904-13917 (2009).

\bibitem{Baravian2005}
C. Baravian, F. Caton, J. Dillet, J. Mougel, ``Steady light transport under flow: characterization of evolving dense random media,'' Phys. Rev. E \textbf{71,} 066603 (2005).

\bibitem{Duncan2008} D. D. Duncan, S.J. Kirkpatrick and R.K. Wang, "Statistics of local speckle contrast,'' J. Opt. Soc. Am. A \textbf{25} (1), 9-15 (2008).

\bibitem{Zakharov2009} P. Zakharov and F. Scheffold, ``Advances in dynamic light scattering techniques'', in: A.A. Kokhanovsky, ed. Light Scattering Reviews 4. Heidelberg: Springer,433Ð468 (2009)

\bibitem{Zakharov2010}
P. Zakharov and F. Scheffold, ``Monitoring spatially heterogeneous dynamics in a drying colloidal thin film,'' Soft Materials, to appear (2010).

\bibitem{briers:review}
J.D. Briers, ``Laser doppler, speckle and related techniques for blood perfusion mapping and imaging,'' Physiological Measurement \textbf{22}, R35-R66 (2001).

\bibitem{dunn:lsi01}
A.K. Dunn, H. Bolay, M.A. Moskowitz, and D.A. Boas, ``Dynamic imaging of cerebral blood flow using laser speckle,''
J. Cereb. Blood Flow Metab. \textbf{21,} 195-201 (2001).

\bibitem{weber:imaging}
B. Weber, C. Burger, M.T. Wyss, G.K. von Schulthess, F. Scheffold, and A. Buck, ``Optical imaging of the spatiotemporal dynamics of cerebral blood flow and oxidative metabolism in the rat barrel cortex,'' Eur. J. Neurosci. \textbf{20}(10), 2664-2670 (2004).

\bibitem{durduran:spat04}
T. Durduran, M.G. Burnett, C. Zhou, G. Yu, D. Furuya, A.G. Yodh, J.A. Detre, and J.H. Greenberg, ``Spatiotemporal quantification of cerebral blood flow during functional activation in rat somatosensory cortex using laser-speckle flowmetry,'' J. Cereb. Blood Flow Metab. \textbf{24,} 518-525 (2004).

\bibitem{dunn:spat05}
A.K. Dunn, A. Devor, A.M. Dale, and D.A. Boas,
``Spatial extent of oxygen metabolism and hemodynamic changes during functional activation of the rat somatosensory cortex,''
NeuroImage \textbf{27}(15), 279-290 (2005).

\bibitem{Yoshimara1986} T. Yoshimara, ``Statistical properties of dynamic speckles," J. Opt. Soc. Am A \textbf{3}, 1032-1054 (1986).

\bibitem{schatzel:noise90}
K. Sch\"atzel, ``Noise on photon correlation data. i. autocorrelation functions,'' Quant. Opt.: J. Eur. Opt. Soc. B \textbf{2}, 287--305 (1990).

\bibitem{Erpelding2008}
M. Erpelding, A. Amon, J.E. Crassous, ``Diffusive wave spectroscopy applied to the spatially resolved deformation of a solid,''
Phys. Rev. E \textbf{78,} 046104 (2008).


\bibitem{WLMPRL07}
N. Willenbacher, C. Oelschlaeger, M. Sch\v{s}pferer, P. Fischer, F. Cardinaux, and F. Scheffold, ``Broad bandwidth optical and mechanical rheometry of wormlike micelle solutions,'' Phys. Rev. Lett. \textbf{99,} 068302 (2007).

\end{thebibliography}
\end{document}